\documentclass[epj]{webofc}
\usepackage[varg]{txfonts}   
\wocname{EPJ Web of Conferences}
\woctitle{CONF12}
\newcommand{\ab}{\bar a}
\newcommand{\ah}{\hat a}
\newcommand{\cO}{{\cal O}}

\newcommand{\nn}{\nonumber}
\newcommand{\eqn}[1]{(\ref{#1})}
\newcommand{\svs}{\vbox{\vskip 4mm}}
\newcommand{\tvs}{\vbox{\vskip 6mm}}
\newcommand{\MSb}{{\overline{\rm MS}}}
\newcommand{\sfrac}[2]{\mbox{$\frac{#1}{#2}$}}

\begin{document}
\selectlanguage{english}
\title{Scheme variations of the QCD coupling}

\author{Diogo Boito\inst{1}
        \and
        Matthias Jamin\inst{2,3}\fnsep\thanks{Speaker. \email{jamin@ifae.es}}
        \and
        Ramon Miravitllas\inst{2}
}

\institute{Instituto de F\'isica de S\~ao Carlos, Universidade de
           S\~ao Paulo, CP 369, 13560-970, S\~ao Carlos, SP, Brazil
\and
           IFAE, BIST, Campus UAB, 08193 Bellaterra (Barcelona) Spain
\and
           ICREA, Pg.~Llu\'\i s Companys 23, 08010 Barcelona, Spain
}

\abstract{%
The Quantum Chromodynamics (QCD) coupling $\alpha_s$ is a central parameter
in the Standard Model of particle physics. However, it depends on theoretical
conventions related to renormalisation and hence is not an observable quantity.
In order to capture this dependence in a transparent way, a novel definition of
the QCD coupling, denoted by $\ah$, is introduced, whose running is explicitly
renormalisation scheme invariant. The remaining renormalisation scheme
dependence is related to transformations of the QCD scale $\Lambda$, and can
be parametrised by a single parameter $C$. Hence, we call $\ah$ the $C$-scheme
coupling. The dependence on $C$ can be exploited to study and improve
perturbative predictions of physical observables. This is demonstrated for the
QCD Adler function and hadronic decays of the $\tau$ lepton.}
\maketitle

\section{Introduction}

A central approach to predictions in Quantum Chromodynamics (QCD) relies on
perturbation theory in the strong coupling $\alpha_s$. However, quarks do
not appear as free particles in nature, they are confined into hadrons, and
the definition of $\alpha_s$ depends on theoretical conventions like the
renormalisation scale or renormalisation scheme. Evidently, physical
observables should not depend on such choices, which is reflected in so-called
renormalisation group equations (RGE's) for scale variations, which measurable
quantities have to satisfy. The situation regarding the scheme dependence is
more involved because the strong coupling can be redefined order by order in
perturbation theory. For this reason, perturbative computations are performed
mainly in convenient schemes such as minimal subtraction (MS)~\cite{hv72} or
modified minimal subtraction ($\MSb$)~\cite{bbdm78}.

This talk reports on a recent work \cite{bjm16}, in which a new definition
of the strong coupling, $\ah$, was introduced. The coupling $\ah$ satisfies two
nice properties: first, its scale evolution, described by the $\beta$-function,
is explicitly scheme invariant. Second, the remaining scheme dependence of $\ah$
can be parametrised by a single parameter $C$. For this reason, henceforth, we
shall refer to $\ah$ as the $C$-scheme coupling, even though $C$ parametrises a
whole class of schemes. Variations of $C$ directly correspond to transformations
of the QCD scale parameter $\Lambda$. Furthermore, it can be demonstrated
that the $C$ dependence of $\ah$ is also governed by the corresponding
$\beta$-function.

After an introduction of the coupling $\ah$ and the discussion of its
properties, we proceed to apply it to phenomenologically relevant cases. One
of the best studied QCD quantities is the two-point vector correlation function
and its derivative, the Adler function~\cite{adl74}, which is related to a
spectral integral over the total cross section for $e^+e^-$ scattering into
hadrons. It also governs theoretical predictions of the inclusive decay rate
of $\tau$ leptons into hadronic final states~\cite{bnp92}. Presently, the
perturbative series for the Adler function is known up to the fourth order in
$\alpha_s$~\cite{bck08}. Exploiting the scheme dependence of the coupling $\ah$
through variations of the parameter $C$, it can be shown how to improve
theoretical predictions for the phenomenological quantities. The use of $\ah$
for the scalar correlator, which is relevant for the prediction of Higgs boson
decay into quarks and for light quark-mass determinations from QCD sum rules,
has been investigated in a related article~\cite{jm16}.

In the past, several other methods have been suggested to optimise perturbative
predictions. In such approaches like BLM \cite{blm83}, or the Principle of
Maximum Conformality (PMC) \cite{bw12,mbw13}, either a scale-setting
prescription is provided to obtain a scheme-independent result, regardless of
the intermediate scheme used for the perturbative calculation (which most often
is $\MSb$). On the other hand, some of these approaches, such as for example
the ``effective charge"~\cite{gru82}, involve a process dependent definition of
the QCD coupling. On the contrary, in the procedure discussed here, one defines
a process-independent class of schemes, parameterised by a single continuous
parameter $C$. Variations of this parameter are then explored in order to
optimise the perturbative series having in mind that we are dealing with
asymptotic expansions. Preferred values of the parameter $C$, however, may
then depend on the quantity under consideration.

\section{The QCD coupling \boldmath{$\hat\alpha_s$}}

To begin, we define the QCD $\beta$-function, which describes the scale
evolution of the coupling $\alpha_s$, as
\begin{equation}
\label{bfun}
-\,Q\,\frac{{\rm d}a_Q}{{\rm d}Q} \,\equiv\, \beta(a_Q) \,=\,
\beta_1\,a_Q^2 + \beta_2\,a_Q^3 + \beta_3\,a_Q^4 + \ldots \,. 
\end{equation}
Here and in the following, we use the abbreviation $a_Q\equiv\alpha_s(Q)/\pi$,
and $Q$ denotes a physically relevant energy scale. The first five
$\beta$-coefficients $\beta_1$ to $\beta_5$ are known analytically
\cite{lrv97,bck16}. (In the conventions employed in this work, they have been
collected in Appendix~A of ref.~\cite{jm16}.) Making use of the RGE~\eqn{bfun}
for $a_Q$, the scale-invariant QCD parameter $\Lambda$ can be defined by
\begin{equation}
\label{Lambda}
\Lambda \,\equiv\, Q\, {\rm e}^{-\frac{1}{\beta_1 a_Q}}
\,[ a_Q ]^{-\frac{\beta_2}{\beta_1^2}}
\exp\Biggl\{\,\int\limits_0^{a_Q}\,\frac{{\rm d}a}{\tilde\beta(a)}\Biggr\}\,,
\end{equation}
with the combination
\begin{equation}
\frac{1}{\tilde\beta(a)} \,\equiv\, \frac{1}{\beta(a)} - \frac{1}{\beta_1 a^2}
+ \frac{\beta_2}{\beta_1^2 a} \,,
\end{equation}
which remains free of singularities in the limit $a\to 0$. Let us consider a
scheme transformation to another coupling $a'$, which assumes the general form
\begin{equation}
\label{ap}
a' \,\equiv\, a + c_1\,a^2 + c_2\,a^3 + c_3\,a^4 + \ldots \,.
\end{equation}
The $\Lambda$-parameter in the transformed scheme, $\Lambda'$, depends only on
$c_1$ and not on the remaining higher-order coefficients. The actual relation
reads~\cite{cg79}
\begin{equation}
\label{Lambdap}
\Lambda' \,=\, \Lambda\,{\rm e}^{c_1/\beta_1} \,.
\end{equation}

\begin{figure}[ht]
\centering
\includegraphics[width=11cm,clip]{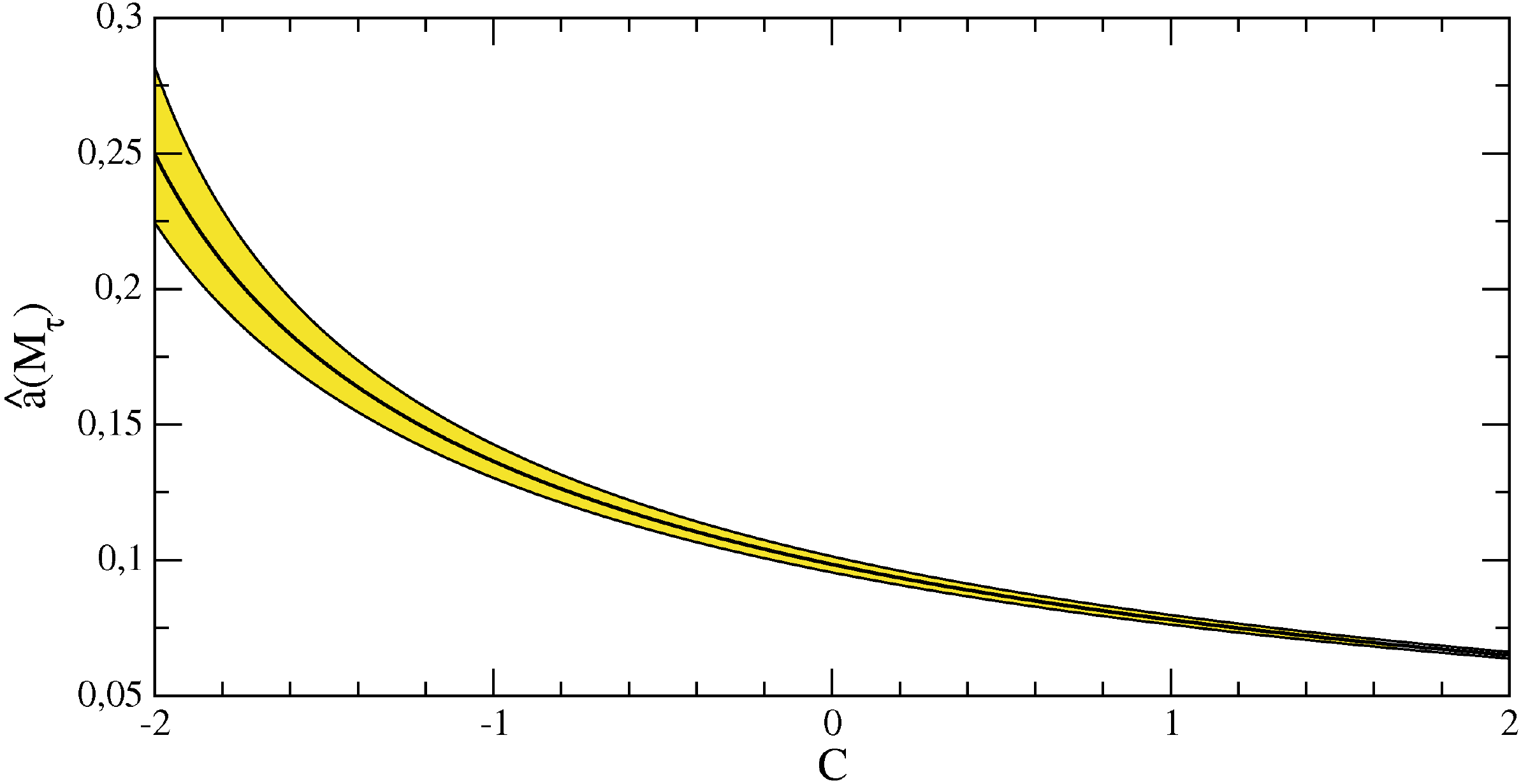}
\caption{The coupling $\ah(M_\tau)$ according to eq.~\eqn{ahat} as a function
of $C$, and for the input value $\alpha_s(M_\tau)=0.316(10)$ in the $\MSb$
scheme.  The yellow band corresponds to the $\alpha_s$ uncertainty.}
\label{fig1}
\end{figure}

The observation that redefinitions of the $\Lambda$-parameter only involve a
single constant motivates an implicit definition of a new coupling $\ah_Q$,
which is scheme invariant except for shifts in $\Lambda$, represented by a
parameter $C$ \cite{bjm16}:
\begin{equation}
\label{ahat}
\frac{1}{\hat a_Q} + \frac{\beta_2}{\beta_1} \ln\hat a_Q -
\frac{\beta_1}{2}\,C \,\equiv\, \beta_1 \ln\frac{Q}{\Lambda} \,=\,
\frac{1}{a_Q} + \frac{\beta_2}{\beta_1}\ln a_Q -
\beta_1 \!\int\limits_0^{a_Q}\, \frac{{\rm d}a}{\tilde\beta(a)} \,.
\end{equation}
In perturbation theory, eq.~\eqn{ahat} should be interpreted in an iterative
sense. Obviously, $\ah_Q$ is a function of $C$ but, for notational simplicity,
this dependence will not be made explicit. It should be remarked that a
combination similar to~\eqn{ahat}, but without the logarithmic term on the
left-hand side, has already been discussed in refs.~\cite{byz92,ben93}. Without
this term, however, an unwelcome logarithm of $a_Q$ remains in the perturbative
relation between the couplings $\ah_Q$ and $a_Q$. This non-analytic contribution
is avoided by the construction of eq.~\eqn{ahat}. In fig.~\ref{fig1}, the
coupling $\ah$ according to eq.~\eqn{ahat} is displayed as a function of $C$.
Since in this work we focus on hadronic $\tau$ decays, our initial $\MSb$ input
is employed as $\alpha_s(M_\tau)=0.316(10)$, which results from the current PDG
average $\alpha_s(M_Z)=0.1181(13)$ \cite{pdg14}. The yellow band corresponds
to the variation within present $\alpha_s$ uncertainties. Below approximately
$C=-2$, the relation between $\ah$ and the $\MSb$ coupling ceases to be
perturbative and breaks down.

As a next step, the $\beta$-function and the corresponding running of the
coupling $\ah$ can be deduced from eq.~\eqn{ahat}. The $\beta$-function is
found to have the rather simple form
\begin{equation}
\label{betahat}
-\,Q\,\frac{{\rm d}\ah_Q}{{\rm d}Q} \,\equiv\, \hat\beta(\ah_Q) \,=\,
\frac{\beta_1 \ah_Q^2}{\left(1 - \sfrac{\beta_2}{\beta_1}\, \ah_Q\right)} \,.
\end{equation}
As is seen explicitly, it only depends on the scheme-invariant $\beta$-function
coefficients $\beta_1$ and $\beta_2$. We also remark that the only non-trivial
zero of $\hat\beta(\ah)$ arises in the case of $\beta_1=0$.
Straightforward integration of the RGE \eqn{betahat} yields
\begin{equation}
\label{ahatrun}
\frac{1}{\ah_Q} \,=\, \frac{1}{\ah_\mu} + \frac{\beta_1}{2}\ln\frac{Q^2}{\mu^2}
- \frac{\beta_2}{\beta_1}\ln\frac{\ah_Q}{\ah_\mu} \,.
\end{equation}
This implicit equation for $\ah_Q$ can either be solved numerically, or
iteratively, to provide a perturbative expansion. The evolution in $C$ can also
be expressed in terms of an RGE. Simply taking the derivative of eq.~\eqn{ahat}
with respect to $C$, one derives the relation
\begin{equation}
\label{CRGE}
\frac{{\rm d}\ah_Q}{{\rm d}C} \,=\, -\,\frac{1}{2}\,\hat\beta(\ah_Q) \,,
\end{equation}
demonstrating that the $C$-``running'' is also governed by the $\beta$-function
$\hat\beta(\ah_Q)$. This explains why the $C$ dependence of $\ah(M_\tau)$
displayed in fig.~\ref{fig1} appears similar to the scale running. They are in
fact equivalent.

The latter observation opens the possibility to arrive at the perturbative
expansion in the $C$-scheme at arbitrary $C$ by first computing the expansion
in $\ah_Q$ at $C=0$, and then employing the evolution equation \eqn{CRGE} to
arrive at an arbitrary $C$. This is completely analogous to the possibility of
reconstructing the scale logarithms from the RGE in the renormalisation scale.
Let us define the abbreviation $\ab_Q\equiv \ah_Q^{C=0}$. Then the relation
between the coupling $a_Q$ and its corresponding $\beta$-function coefficients
in an arbitrary scheme, for example the $\MSb$ scheme, and $\ab_Q$ is found to
be
\begin{eqnarray}
\label{afunab}
a_Q \,=\, \ab_Q \!\!&+&\!\! \biggl( \frac{\beta_3}{\beta_1} -
\frac{\beta_2^2}{\beta_1^2} \biggr) \,\ab_Q^3 + \biggl(
\frac{\beta_4}{2\beta_1} - \frac{\beta_2^3}{2\beta_1^3} \biggr) \,\ab_Q^4
\nn \\
\tvs
&+&\!\! \biggl( \frac{\beta_5}{3\beta_1} - \frac{\beta_2\beta_4}{6\beta_1^2} +
\frac{5\beta_3^2}{3\beta_1^2} - \frac{3\beta_2^2\beta_3}{\beta_1^3} +
\frac{7\beta_2^4}{6\beta_1^4} \biggr) \,\ab_Q^5 + {\cal O}(\ab_Q^6) \,.
\end{eqnarray}
The successive relation between the coupling $\ah_Q$ at arbitrary $C$ and
$\ab_Q$ which only contains the scheme-invariant coefficients $\beta_1$ and
$\beta_2$, can then be derived from the RGE~\eqn{CRGE}. It is found to take
the form
\begin{eqnarray}
\label{abfunah}
\ab_Q \,=\, \ah_Q \!\!\!&+&\!\!\! \frac{\beta_1}{2}\,C\,\ah_Q^2 + \biggl(
\frac{\beta_2}{2}\,C + \frac{\beta_1^2}{4}\,C^2 \biggr) \,\ah_Q^3
+ \biggl( \frac{\beta_2^2}{2\beta_1}\,C + \frac{5\beta_1\beta_2}{8}\,C^2
+ \frac{\beta_1^3}{8}\,C^3 \biggr) \,\ah_Q^4 \nn \\
\tvs 
&+&\!\!\! \biggl( \frac{\beta_2^3}{2\beta_1^2}\,C + \frac{9\beta_2^2}{8}\,C^2 +
\frac{13\beta_1^2\beta_2}{24}\,C^3 + \frac{\beta_1^4}{16}\,C^4 \biggr) \,\ah_Q^5
+ {\cal O}(\ah_Q^6) \,.
\end{eqnarray}

Inserting now eq.~\eqn{abfunah} into \eqn{afunab}, the perturbative relations
between the coupling $\ah$ and $a$ in a particular scheme can straightforwardly
be deduced. Taking $a$ as well as the respective $\beta$-function coefficients
in the $\MSb$ scheme, and for three quark flavors, $N_f=3$, the expansions
read, 
\begin{eqnarray}
\label{aofahat}
a(\ah) \!\!\!&=&\!\!\! \ah + \frac{9}{4}\,C\,\ah^2 + \biggl( \frac{3397}{2592}
+ 4 C + \frac{81}{16}\,C^2 \biggr) \,\ah^3 + \biggl( \frac{741103}{186624} 
+ \frac{18383}{1152}\,C + \frac{45}{2}\,C^2 \nn \\
\tvs
&& +\, \frac{729}{64}\,C^3 + \frac{445}{144}\zeta_3 \biggr) \,\ah^4 + \biggl(
\frac{1142666849}{80621568} + \frac{1329359}{20736}\,C + \frac{28623 }{256}\,C^2
+ \frac{351}{4}\,C^3 \nn \\
\tvs
&& +\, \frac{6561}{256}\,C^4 + \frac{445}{16}\zeta_3 C +
\frac{10375693}{373248}\zeta_3 - \frac{1335}{256}\zeta_4
- \frac{534385}{20736}\zeta_5 \biggr) \,\ah^5 + \cO(\ah^6) \,,
\end{eqnarray}
and
\begin{eqnarray}
\label{ahatofa}
\ah(a) \!\!\!&=&\!\!\! a - \frac{9}{4}\,C\,a^2 - \biggl( \frac{3397}{2592} +
4 C - \frac{81}{16}\,C^2 \biggr) \,a^3 - \biggl( \frac{741103}{186624} 
+ \frac{233}{192}\,C - \frac{45}{2}\,C^2 \nn \\
\tvs
&& +\, \frac{729}{64}\,C^3 + \frac{445}{144}\zeta_3 \biggr) \,a^4 -
\biggl(\frac{727240925}{80621568}  - \frac{869039}{41472}\,C -
\frac{26673}{512}\,C^2 + \frac{351}{4}\,C^3 \nn \\
\tvs
&& -\, \frac{6561}{256}\,C^4 - \frac{445}{32}\zeta_3 C 
+ \frac{10375693}{373248}\zeta_3 - \frac{1335}{256}\zeta_4
- \frac{534385}{20736}\zeta_5 \biggr) \,a^5 + \cO(a^6) \,,
\end{eqnarray}
where $\zeta_i\equiv \zeta(i)$ stands for the Riemann $\zeta$-function.
Eqs.~\eqn{aofahat} and \eqn{ahatofa} have originally been presented in
ref.~\cite{bjm16}.

\section{The Adler function}

As our first application of the coupling $\hat\alpha_s$, the perturbative
series of the Adler function $D(a_Q)$~\cite{adl74,bck08} shall be investigated.
To this end, it is convenient to define the reduced Adler function $\hat D(a_Q)$
as
\begin{equation}
\label{Dhat}
4\pi^2 D(a_Q) - 1 \,\equiv\, \hat D(a_Q) \,=\, \sum_{n=1}^\infty c_{n,1} a_Q^n
\,=\, a_Q + 1.640\,a_Q^2 + 6.371\,a_Q^3 + 49.08\,a_Q^4 +\ldots \,.
\end{equation}
We adopt the notation of ref.~\cite{bj08}, with numerical coefficients at
$N_f=3$ and in the $\MSb$ scheme. The renormalisation scale logarithms
$\ln(Q/\mu)$ appearing in the Adler function have been resummed with the scale
choice $\mu=Q$.

\begin{figure}[ht]
\centering
\includegraphics[width=11cm,clip]{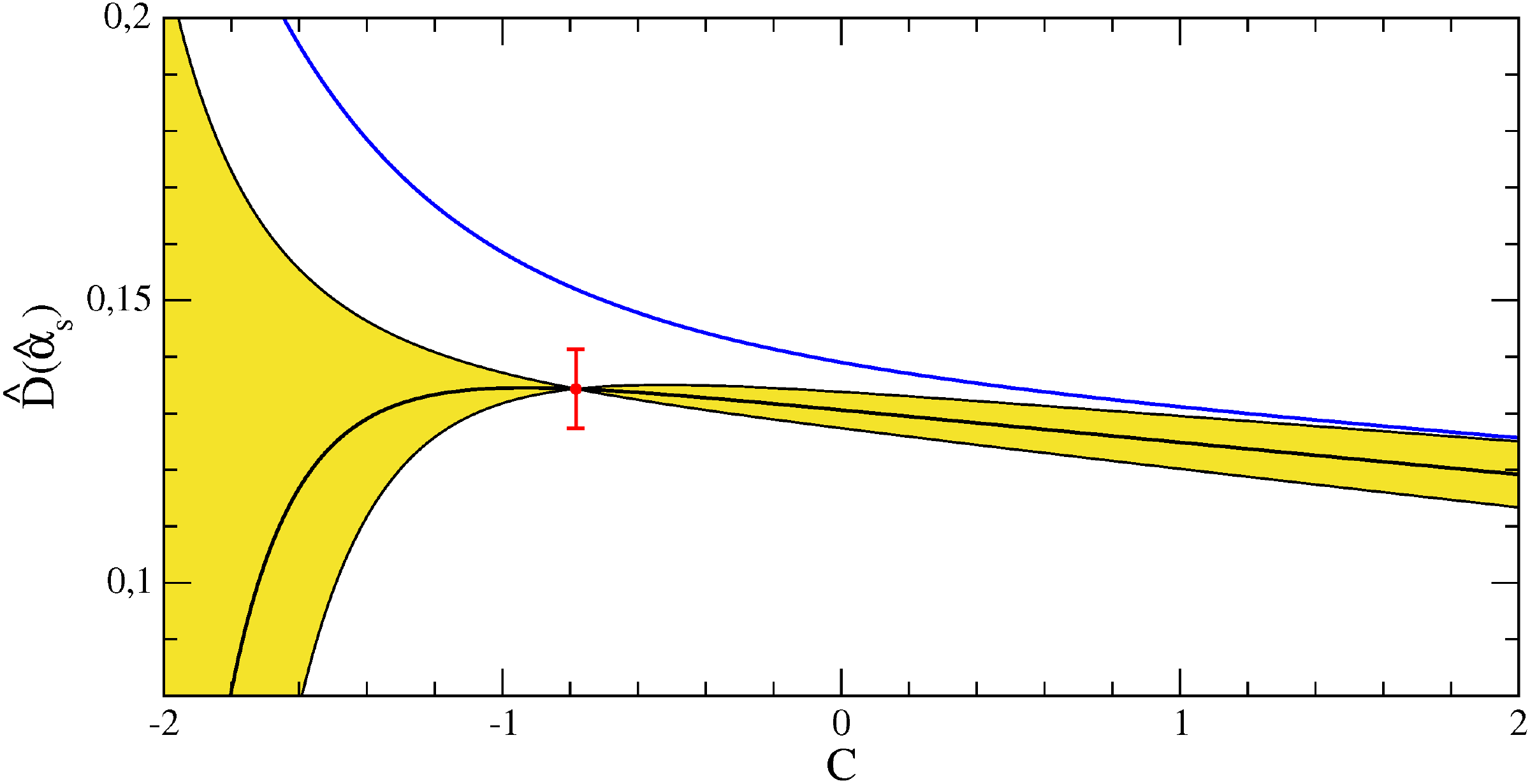}
\caption{$\hat D(\ah_{M_\tau})$ of eq.~\eqn{Dhatah} as a function of $C$. The
yellow band arises from either removing or doubling the fifth-order term. In
the red dot, the $\cO(\ah^5)$ vanishes, and $\cO(\ah^4)$ is taken as the
uncertainty. For further explanation, see the text.}
\label{fig2}
\end{figure}

Employing the relation \eqn{aofahat}, the expansion \eqn{Dhat} for $\hat D$
can be rewritten in terms of the $C$-scheme coupling $\ah_Q$, resulting in
\begin{eqnarray}
\label{Dhatah}
\hat D(\ah_Q) \!\!\!&=&\!\!\! \ah_Q + (1.640 + 2.25 C)\,\ah_Q^2 
+ (7.682 + 11.38 C + 5.063 C^2)\,\ah_Q^3 \nn \\
\svs
&&\!\!\!  +\, (61.06 + 72.08 C + 47.40 C^2 + 11.39 C^3)\,\ah_Q^4 + \ldots \,.
\end{eqnarray}
A graphical representation of eq.~\eqn{Dhatah} is provided in Fig.~\ref{fig2},
where $\hat D(\ah_{M_\tau})$ is displayed as a function of the scheme parameter
$C$. Here, the yellow band represents an error estimate from the fifth-order
contribution. The required coefficient was taken to be $c_{5,1}=283$, according
to an estimate deduced in ref.~\cite{bj08}. The yellow band then corresponds to
either removing or doubling the $\cO(\ah^5)$ term. Generally, it is observed
that around $C\!\approx\!-1$, a region of stability with respect to
$C$-variations emerges. For comparison, the blue line corresponds to employing
$c_{5,1}=566$ and in addition doubling the $\cO(\ah^5)$ correction. Then, no
stability is found which seems to indicate that such large values of $c_{5,1}$
are disfavoured. In the red dot, where $C=-0.783$, the $\cO(\ah^5)$ vanishes,
and the $\cO(\ah^4)$ correction has been employed as a conservative uncertainty,
which is the last included non-vanishing term, in view of the asymptotic nature
of the series. Numerically, it reads
\begin{equation}
\label{Dhoa5zero}
\hat D(\ah_{M_\tau},C=-0.783) \,=\, 0.1343 \pm 0.0070 \pm 0.0067 \,,
\end{equation}
where the second error originates from the uncertainty in $\alpha_s(M_\tau)$.
The result \eqn{Dhoa5zero} can be compared to the direct $\MSb$ prediction
\eqn{Dhat}, which corresponds to
\begin{equation}
\label{DhatMSb}
\hat D(a_{M_\tau}) \,=\, 0.1316 \pm 0.0029 \pm 0.0060 \,.
\end{equation}
Here, the first error is obtained by removing or doubling $c_{5,1}$, and the
second error again reflects the parametric $\alpha_s$ uncertainty.

A final comparison of \eqn{Dhoa5zero} and \eqn{DhatMSb} may be performed with
the Adler function model that was put forward in ref.~\cite{bj08}, and which is
based on general knowledge of the renormalon structure for the Borel transform
of $\hat D(a_Q)$. Within this model, one obtains
\begin{equation}
\label{DhatBM}
\hat D(a_{M_\tau}) \,=\, 0.1354 \pm 0.0127 \pm 0.0058 \,.
\end{equation}
In this case, the first uncertainty results from estimates of the perturbative
ambiguity that arises from the renormalon singularities. It is observed that
this uncertainty is substantially bigger than the one of \eqn{DhatMSb} and
still larger than the one of \eqn{Dhoa5zero}. Therefore, we conclude that the
higher-order uncertainty of \eqn{DhatMSb} appears to be underestimated, while
eq.~\eqn{Dhoa5zero} seems to provide a more realistic account of the resummed
series. Interestingly enough, also its central value is closer to the Borel
model result.

\begin{figure}[ht]
\centering
\includegraphics[width=11cm,clip]{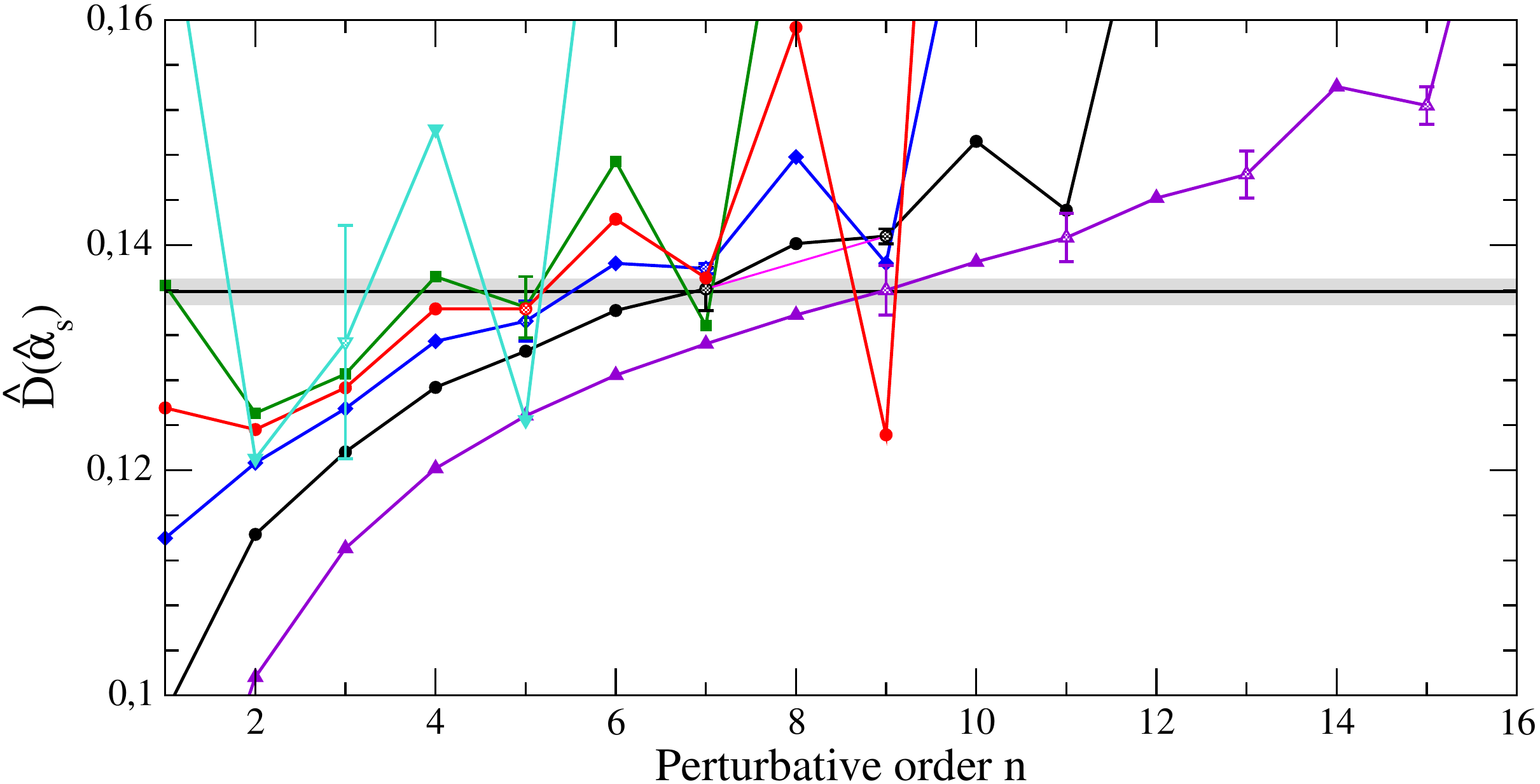}
\caption{Partial sums up to order $n$ of the perturbative series for
$\hat D(\hat\alpha_s)$ for different values of $C$: $C=1$ (violet); $C=0$
(black); $C=-0.5$ (blue); $C=-0.783$ (red); $C=-1$ (green); $C=-1.5$ (cyan).
The grey band corresponds to the Borel sum according to the central model of
ref.~\cite{bj08}. For further details and discussion see the text.}
\label{fig3}
\end{figure}

To conclude this section, let us investigate the behaviour of the perturbative
series for the Adler function $\hat D(\hat\alpha_s)$ order by order. To this
end, in figure~\ref{fig3}, we display the partial sums up to order $n$ for
different values of the scheme parameter $C$. The colour coding is as follows:
C=1 (violet); C=0 (black); C=-0.5 (blue); C=-0.783 (red); C=-1 (green); C=-1.5
(cyan). Perturbative orders higher than 5 have been taken according to the
central model for the Borel-transformed Adler function of ref.~\cite{bj08},
expressed in the coupling $\hat\alpha_s$. Furthermore, the grey band corresponds
to the Borel sum also according to this model.

We make the following observations: at low orders, the series is dominated by
the infrared (IR) renormalons which for C=0 yield a fixed sign series. (At
$C<0$, as can be seen from the figure, this need no longer be the case.) At
larger order the ultraviolet (UV) renormalons, being sign alternating, take
over, and ultimately, the leading UV renormalon at $u=-1$ dominates. If $C$
becomes smaller and smaller, the dominance of UV renormalons starts at lower
and lower orders and the general behaviour of the series becomes worse. As the
shaded symbols with error bars, we denote small terms in the series. The error
bar indicates the size of the respective term. For a strictly sign-alternating
series, the smallest term would correspond to the closest approach of the series
to the full result.  Because we have fixed-sign and alternating-sign components
in our series, we find terms accidentally small, and sometimes smaller than
the small terms which are close to the Borel sum. In addition, for smaller $C$,
the smallest term appears at lower orders, but also its size increases (and so
would the corresponding error estimate). In the future, we plan to exploit the
$C$ dependence of the Adler function series in $\hat\alpha_s$, in order to tune
the smallest term to the presently available number of orders and to obtain
more reliable error estimates.

\section{The total tau hadronic width}

We now turn our attention to the perturbative expansion for the total $\tau$
hadronic width. The central observable is the ratio $R_\tau$ for the total
hadronic branching fraction to the electron branching fraction. It can be
expressed as
\begin{equation}
R_\tau \,=\, 3\, S_{\rm EW} (|V_{ud}|^2 + |V_{us}|^2)\, ( 1 + \delta^{(0)}
+ \cdots) \,,
\end{equation}
where $S_{\rm EW}$ is an electroweak correction and $V_{ud}$ as well as
$V_{us}$ CKM matrix elements. The perturbative QCD contribution is contained in
$\delta^{(0)}$ and the ellipsis indicate further small subleading corrections.
(See refs.~\cite{bnp92,bj08} for  details.) For $\delta^{(0)}$ a complication
arises, because it is calculated from a contour integral in the complex energy
plane. On the other hand, we seek to resum the scale logarithms $\ln(Q/\mu)$,
and the perturbative prediction depends on whether those logs are resummed
after or before performing the contour integration. The first choice is called
fixed-order perturbation theory (FOPT) and the second contour-improved 
perturbation theory (CIPT) \cite{dp92}.

In FOPT, the perturbative series of $\delta^{(0)}(a_Q)$ in terms of the $\MSb$
coupling $a_Q$ reads \cite{bck08,bj08}
\begin{equation}
\label{del0}
\delta_{\rm FO}^{(0)}(a_Q) \,=\,
a_Q + 5.202\,a_Q^2 + 26.37\,a_Q^3 + 127.1\,a_Q^4 +\ldots \,.
\end{equation}
On the other hand, in the $C$-scheme coupling $\ah_Q$, the expansion for
$\delta^{(0)}(\ah_Q)$ is given by
\begin{eqnarray}
\label{del0ah}
\delta_{\rm FO}^{(0)}(\ah_Q) \!\!\!&=&\!\!\! \ah_Q + (5.202 + 2.25 C)\,\ah_Q^2
+ (27.68 + 27.41 C + 5.063 C^2)\,\ah_Q^3 \nn \\
\svs
&&\!\!\! +\, (148.4 + 235.5 C + 101.5 C^2 + 11.39 C^3)\,\ah_Q^4 + \ldots \,.
\end{eqnarray}

\begin{figure}[ht]
\centering
\includegraphics[width=11cm,clip]{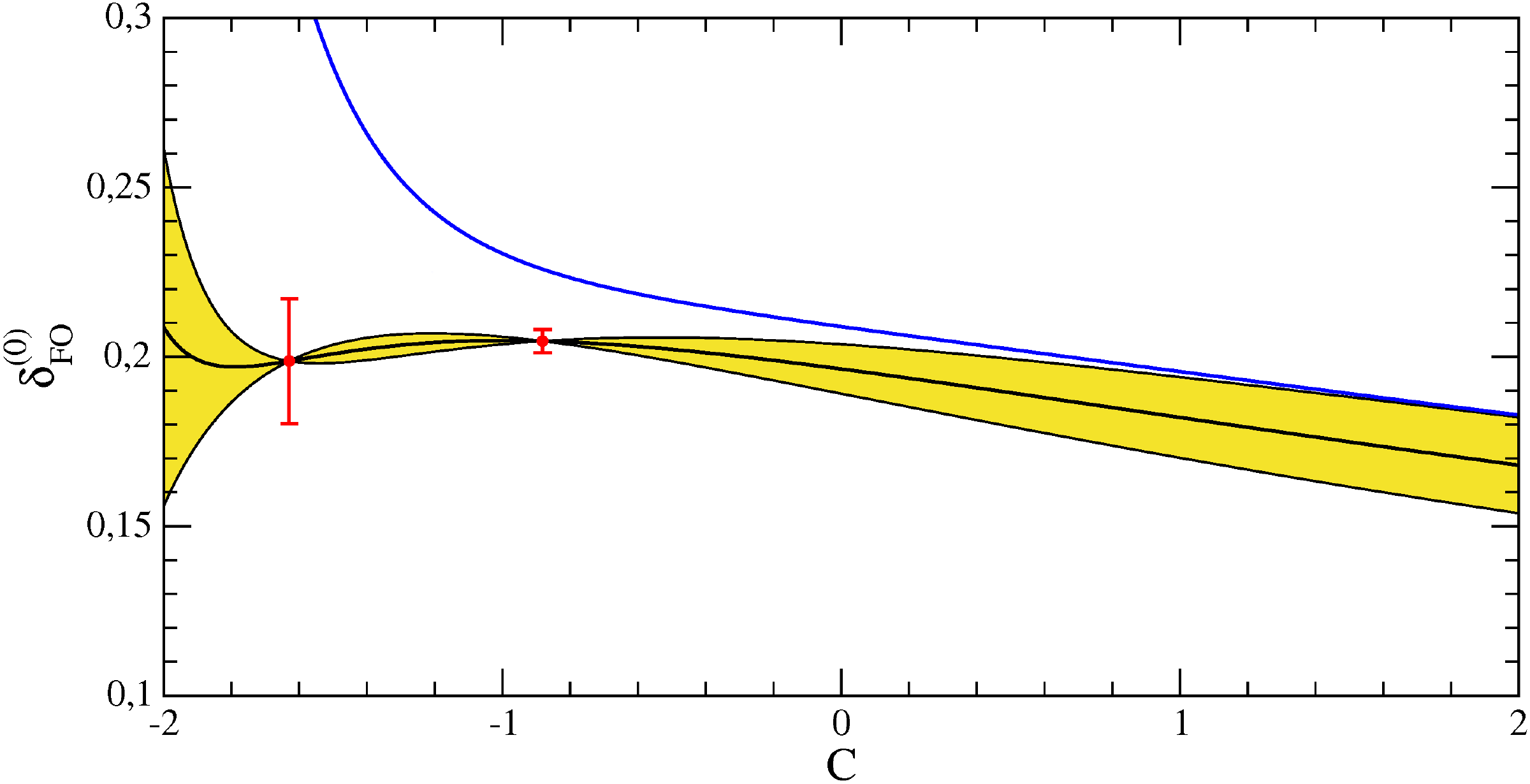}
\caption{$\delta_{\rm FO}^{(0)}(\ah_Q)$ of eq.~\eqn{del0ah} as a function of
$C$. The yellow band arises from either removing or doubling the fifth-order
term. In the red dots, the $\cO(\ah^5)$ vanishes, and $\cO(\ah^4)$ is taken as
the uncertainty. For further explanation, see the text.}
\label{fig4}
\end{figure}

In Fig.~\ref{fig4}, $\delta_{\rm FO}^{(0)}(\ah_Q)$ is displayed as a function
of $C$. Assuming $c_{5,1}=283$, the yellow band again corresponds to removing
or doubling the $\cO(\ah^5)$ term. As for $\hat D(\ah)$, a nice plateau is
found for $C\approx -1$. Taking $c_{5,1}=566$ and then doubling the $\cO(\ah^5)$
results in the blue curve which does not show stability. Hence, this scenario
again is disfavoured. In the red dots, which are located at $C=-0.882$ and
$C=-1.629$, the $\cO(\ah^5)$ correction vanishes, and the $\cO(\ah^4)$ term is
taken as the uncertainty. The point to the right has a substantially smaller
error, and yields
\begin{equation}
\label{del0oa5zero}
\delta_{\rm FO}^{(0)}(\ah_{M_\tau},C=-0.882) \,=\,
0.2047 \pm 0.0034 \pm 0.0133 \,.
\end{equation}
Like for the Adler function, the second error covers the parametric uncertainty
for $\alpha_s(M_\tau)$. In this case, the direct $\MSb$ prediction of
eq.~\eqn{del0} is found to be
\begin{equation}
\label{del0MSb}
\delta_{\rm FO}^{(0)}(a_{M_\tau}) \,=\, 0.1991 \pm 0.0061 \pm 0.0119 \,.
\end{equation}
This value is somewhat lower, but within $1\,\sigma$ of the higher-order
uncertainty. Comparing, on the other hand, to the Borel model (BM) result
of \cite{bj08}, which is given by
\begin{equation}
\label{del0BM}
\delta_{\rm BM}^{(0)}(a_{M_\tau}) \,=\, 0.2047 \pm 0.0029 \pm 0.0130 \,,
\end{equation}
it is found that \eqn{del0oa5zero} and \eqn{del0BM} are surprisingly similar.
In both cases, the $\alpha_s$ uncertainty is substantially larger than the
higher-order one  -- especially given the recent increase in the $\alpha_s$
uncertainty provided by the PDG \cite{pdg14} -- which underlines the good
potential of $\alpha_s$ extractions from hadronic $\tau$ decays.

\begin{figure}[ht]
\centering
\includegraphics[width=11cm,clip]{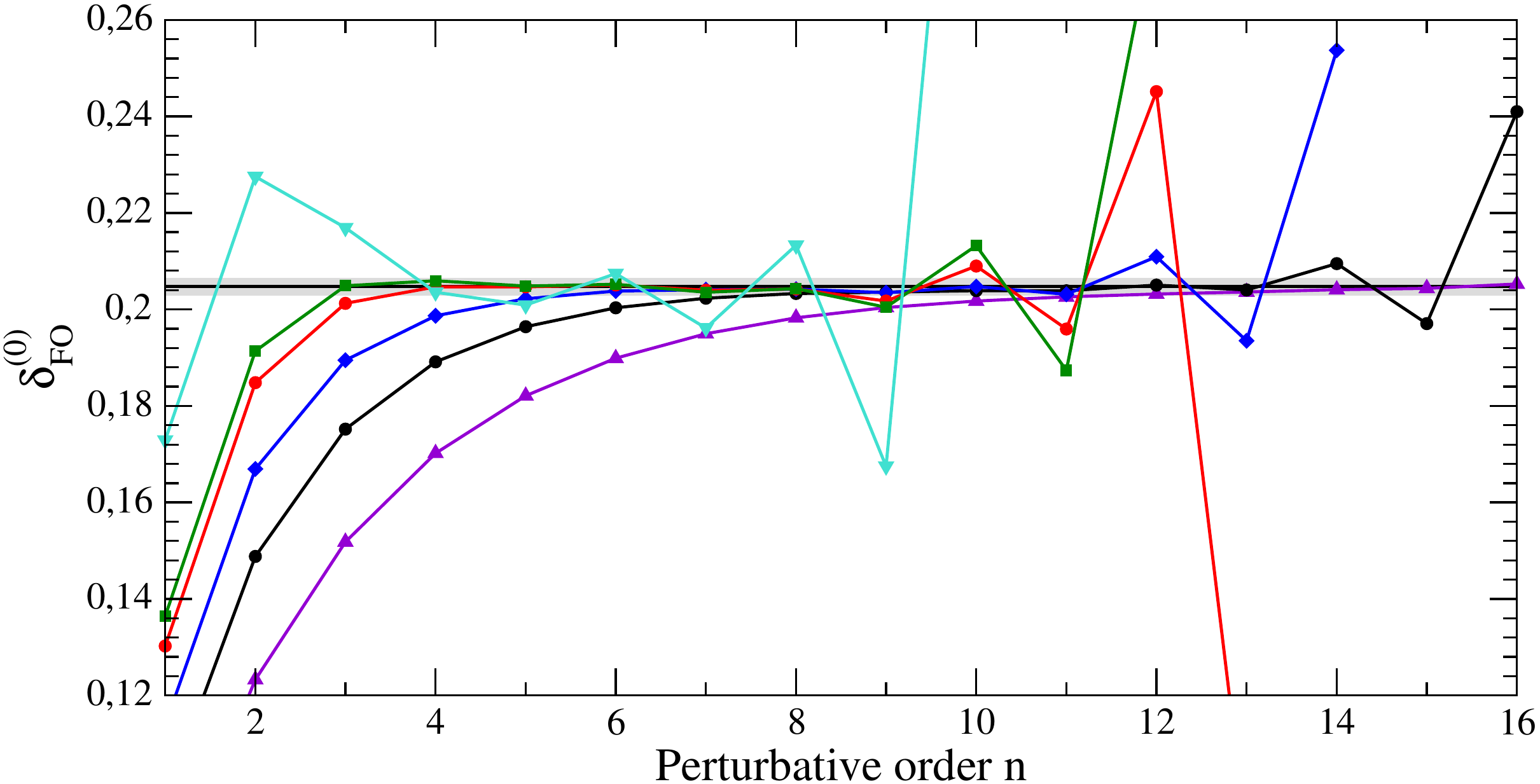}
\caption{Partial sums up to order $n$ of the perturbative series for
$\delta_{\rm FO}^{(0)}(\ah_{M_\tau})$ for different values of $C$: $C=1$
(violet); $C=0$ (black); $C=-0.5$ (blue); $C=-0.882$ (red); $C=-1$ (green);
$C=-1.5$ (cyan). The grey band corresponds to the Borel sum according to the
central model of ref.~\cite{bj08}. For further discussion see the text.}
\label{fig5}
\end{figure}

In figure~\ref{fig5}, we again provide partial sums up to order $n$ of the
perturbative series for $\delta_{\rm FO}^{(0)}(\ah_{M_\tau})$ at different
values of $C$. The colour coding is equivalent to figure~\ref{fig3}: $C=1$
(violet); $C=0$ (black); $C=-0.5$ (blue); $C=-0.882$ (red); $C=-1$ (green);
$C=-1.5$ (cyan). The $C=0$ series is close to the corresponding one in the
$\MSb$ scheme. Although the general perturbative behaviour is similar to the
Adler function, one observes that for $\delta_{\rm FO}^{(0)}(\ah_{M_\tau})$
it is substantially smoother. This is due to a suppression of the leading IR
renormalon pole related to the gluon condensate. For larger $C$, the onset of
the asymptotic behaviour is delayed, however the series needs more terms to
approach the resummed result. For smaller $C$, and this is the case in
particular for $C\approx -1$, the onset of the asymptotic behaviour is earlier,
but also the series requires less terms to come close to the ``true'' value.
In the region around $C=-0.882$ (red and green curves), a satisfactory approach
to the Borel sum is achieved with just the four analytically available orders.
If $C$ is still taken to be smaller, the behaviour of the series deteriorates,
and error estimates from small terms become larger.

\begin{figure}[ht]
\centering
\includegraphics[width=11cm,clip]{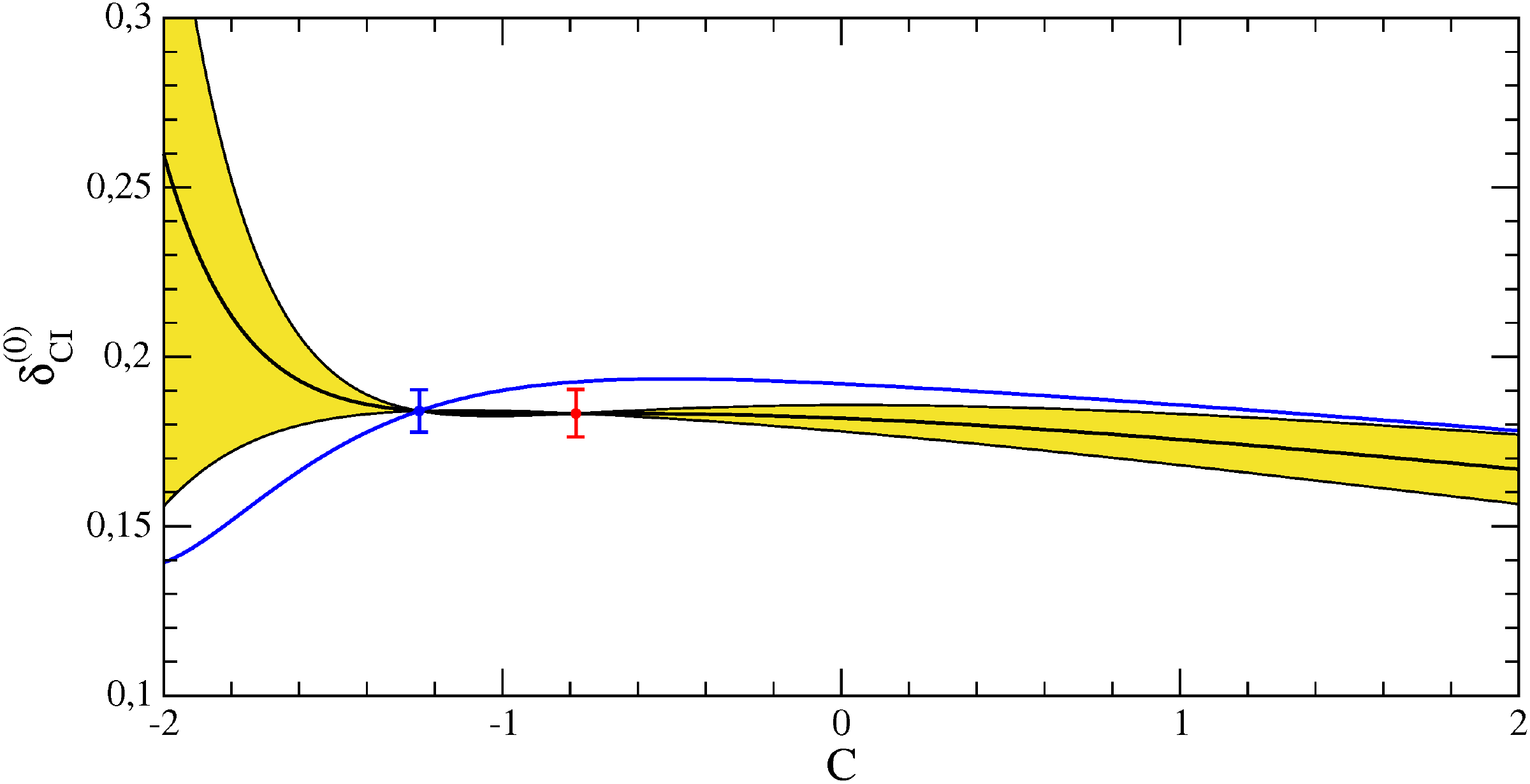}
\caption{$\delta_{\rm CI}^{(0)}(\ah_Q)$ as a function of $C$. The yellow band
arises from either removing or doubling the fifth-order term. In the red and
blue dots, the $\cO(\ah^5)$ vanishes, and $\cO(\ah^4)$ is taken as the
uncertainty. For further explanation, see the text.}
\label{fig6}
\end{figure}

In CIPT, contour integrals over the running coupling, eq.~(\ref{ahatrun}),
have to be computed, and hence the result cannot be given in analytical form.
Graphically, as a function of $C$, $\delta_{\rm CI}^{(0)}(a_{M_\tau})$ is
displayed in Fig.~\ref{fig6}. The general behavior is very similar to FOPT,
with the exception that now also for $c_{5,1}=566$ a zero of the $\cO(\ah^5)$
term is found. This time, both zeros  have similar uncertainties, and employing
the point with smaller error (in blue) results in
\begin{equation}
\label{del0CIoa5z}
\delta_{\rm CI}^{(0)}(\ah_{M_\tau},C=-1.246) \,=\,
0.1840 \pm 0.0062 \pm 0.0084 \,.
\end{equation}
As has been discussed many times in the past (see e.g.~\cite{bj08}) the CIPT
result lies substantially below the FOPT prediction, especially the $C$-scheme
ones, and the Borel model. On the other hand, the parametric $\alpha_s$
uncertainty in CIPT turns out to be smaller.

\section{Conclusions}

In ref.~\cite{bjm16}, see eq.~\eqn{ahat}, we have defined a class of QCD
couplings $\ah_Q$, such that the scale running is explicitly scheme invariant,
and scheme changes can be parameterised by a single constant $C$. For this
reason, we have termed $\ah_Q$ the $C$-scheme coupling. Scheme transformations
correspond to shifts in the QCD scale $\Lambda$. It is furthermore seen that $C$
changes are also governed by the $\beta$-function $\hat\beta(\ah_Q)$ and hence
scale and scheme transformations are equivalent.

We have applied the coupling $\ah_Q$ to investigations of the perturbative
series of the reduced Adler function $\hat D$. Our central result is given in
eq.~\eqn{Dhoa5zero}. Its higher-order uncertainty turned out larger than the
corresponding $\MSb$ prediction \eqn{DhatMSb}, but we consider \eqn{Dhoa5zero}
to be more conservative and realistic.

We also studied the perturbative expansion of the $\tau$ hadronic width,
employing the coupling $\ah_Q$. In this case our central prediction in FOPT
is given in eq.~\eqn{del0oa5zero}. Surprisingly, the result \eqn{del0oa5zero}
is found very close to the prediction \eqn{del0BM} of the central Borel model
developed in ref.~\cite{bj08}, hence providing some support for this approach.

The disparity between FOPT and CIPT predictions for $\delta^{(0)}$ is not
resolved by the $C$-scheme.  As is seen from eq.~\eqn{del0CIoa5z} and
figure~\ref{fig6}, the CIPT result turns out substantially lower (as is
the case for the $\MSb$ prediction). This suggests to return to detailed
investigations of Borel models, this time expressed in the $C$-scheme coupling
$\ah$, in order to investigate the scheme dependence of such models. This could
result in an improved extraction of $\alpha_s$ from hadronic decays of the
$\tau$ lepton.

\vspace{3mm}
\noindent
{\bf Acknowledgments}

\noindent
Helpful discussions with Martin~Beneke are gratefully acknowledged.
The work of MJ and RM has been supported in part by MINECO Grant number
CICYT-FEDER-FPA2014-55613-P, by the Severo Ochoa excellence program of MINECO,
Grant SO-2012-0234, and Secretaria d'Universitats i Recerca del Departament
d'Economia i Coneixement de la Generalitat de Catalunya under Grant 2014 SGR
1450. DB's is supported by the S\~ao Paulo Research Foundation (FAPESP) grant
15/20689-9, and by CNPq grant 305431/2015-3.

\end{document}